\documentclass[epsf]{article}
\usepackage{graphicx}
\begin{document}
\baselineskip7mm
\title{Recollapsing Bianchi I brane worlds}
\author{A. V. Toporensky$^{\dagger}$ and P.V. Tretyakov$^{\ddagger}$}
\date{}
\maketitle
\hspace{8mm}{\em Sternberg Astronomical Institute,
Universitetsky prospekt, 13,  Moscow 119899, Russia}

\begin{abstract}
We investigate the possibility for a flat Bianchi I
brene Universe to recollaps due to the presence of a negative
"dark radiation" and an anisotropic stress in the form
of homogeneous magnetic field, localized on the brane.
\end{abstract}

$^{\ddagger}$ Electronic mail: tpv@xray.sai.msu.ru\\
$^{\dagger}$ Electronic mail: lesha@sai.msu.ru\\

The cosmology of a brane Universe, intensively studied during last
several years, has a variety of possible regimes. Some of these
are impossible in the standard cosmological scenario. 
The quadratic stress-energy 
term in the right-hand side of the effective equation of motion 
\cite{Bin, Cline}
 can result in a nonstandard effective
equation of state of a brane matter
$p=(\gamma-1) \mu$ with 
$2 < \gamma \le 4$, which is forbidden 
in the standard cosmology. Another important
feature of brane models is the presence of projection of 5-dimensional
Weil tensor onto the brane in the equations of motion.
This projection can be decomposed into scalar,
vector and tensor parts \cite{Roy}. 
The scalar part, the only compatible with FRW metric on a brane, 
also called "dark radiation". It
can be either positive or negative, violating in the latter case the
positive energy condition. Both these features (the quadratic corrections
and the presence of the "dark radiation") 
can significantly modify the cosmological
dynamics even in the simplest case of an FRW brane. One of the most
important modifications alters future asymptotic,
because
a large enough negative "dark radiation" can cause a recollaps of a flat
brane Universe \cite{Santos}. 

 The dynamics of a Bianchi I brane  also have been studied in detail. 
In general, as we lift the assumption of spatial
isotropy, possible sources of an anisotropic
stress should be taken into account. However, 
the tensor part of the Weil tensor projection 
onto a brane, being a nonlocal anisotropic stress,
has no evolution equation \cite{Roy}. When some form of 
 nonlocal anysotropic stress is put into equation of 
motion "by hand", the resulting dynamics
 significantly depends on the ansatz chosen\cite{Roy2, Langlois}. 
On the other hand, in the absence of the
Weil tensor, a Bianchi I brane with a perfect fluid allows a complete
description \cite{C-S1, C-S2}. 

The next step in understanding details of brane cosmology have been done
in \cite{B-H}, when a Bianchi I brane with a homogeneous
magnetic field is studied. Being a source of anisotropic stress, the magnetic
field is localized on the brane. The nonlocal anisotropic stress
was chosen to be zero (so that the equations of motion can be written in
a closed form), the "dark energy" ${\cal U}$, however, can not vanish 
identically,
because the magnetic field enters as a source term into the evolution
equation for ${\cal U}$ (see below). The resulting dynamical system 
has many stable
points and a very complex behavior. Due to this complexity some
important issues of magnetic brane evolution remain unclear even
after a careful stable point analysis of \cite{B-H}  
and require numerical studies.

Denoting the combination of the gravitational constant $G$ and
the brane tension $\lambda$ as a new parameter
 $\chi = (8 \pi G \lambda)^{-1}$,
one can write the set of equations which governs the evolution
of a brane magnetic Universe in the form
(see a detailed derivation and a description 
of the reference frame used in \cite{B-H})

\begin{equation}
\Theta^2=3\rho+\frac{3}{2}\sigma^2+\frac{3}{2}h^2+\chi(\frac{3}{2}\rho^2+
\frac{3}{2}\rho h - \frac{3}{4}h^4) + 18 \chi {\cal U}
\end {equation}

\begin{equation}
\dot h=-\frac{2}{3}\Theta h - 2 \sigma_+h
\end{equation}

\begin{equation}
\dot \rho=-\gamma\Theta\rho
\end{equation}

\begin{equation}
\dot \Theta=-\frac{1}{3}\Theta^2-\sigma^2+\frac{1}{2}
[(3\gamma-2)\rho + h^2]-\frac{\chi}{8}[4(3\gamma-1)\rho^2+
(6\gamma+4)\rho h^2 + 3 h^4]-6\chi {\cal U}
\end{equation}

\begin{equation}
\dot {\cal U}=-\frac{4}{3}\Theta {\cal U} -\frac{h^2}{6}
[\theta h^2 +\frac{5}{3}\sigma_+ h^2 -3\gamma\sigma_+ \rho]
\end{equation}

\begin{equation}
\dot \sigma_+=-\Theta\sigma_+ -\sigma^2_{12}-\sigma^2_{13}+
\frac{1}{3}h^2(1-\frac{\chi}{2}[(3\gamma-2)\rho +\frac{4}{3}h^2])
\end{equation}

\begin{equation}
\dot \sigma_-=-\Theta\sigma_- -\frac{1}{\sqrt{3}}
(\sigma^2_{12}-\sigma^2_{13})
\end{equation}

\begin{equation}
\dot \sigma_{23}=-\Theta\sigma_{23}-2\sigma_{12}\sigma_{13}
\end{equation}

\begin{equation}
\dot \sigma_{12}=-\Theta\sigma_{12}+(3\sigma_+ + \sqrt{3}\sigma_-)
\sigma_{12}+\sigma_{13}\sigma_{23}
\end{equation}

\begin{equation}
\dot \sigma_{13}=-\Theta\sigma_{13}+(3\sigma_+ - \sqrt{3}\sigma_-)
\sigma_{13} + \sigma_{12}\sigma_{23}
\end{equation}

Here $\Theta=3H$ is the volume expansion rate, $h$ is the 
nonvanishing component of the magnetic field
(we use the reference frame in which two other components of
the magnetic vector vanish), $\sigma_{\mu \nu}$ is the shear tensor.
The variables $\sigma_+$ and $\sigma_-$ are connected with the shear
components as follows
$$
(\sigma_{11}, \sigma_{22}, \sigma_{33}) = (-2\sigma_+, \sigma_+ + 
\sqrt{3}\sigma_-, \sigma_+ - \sqrt{3}\sigma_-).
$$ 

Moreover, it is possible to define new shear variables
 $\sigma_A$, $\sigma_B$, $\sigma_C$  using the transformation

\begin{equation}
\sigma_{12}+i\sigma_{13}=\sigma_A e^{i\varphi}
\end{equation}

\begin{equation}
\sqrt{3}\sigma_- + i\sigma_{23}=(\sigma_B+i\sigma_C)e^{2i\varphi}
\end{equation}

Due to symmetries of our dynamical system, the equation of motion
for the phase $\varphi$ has the form
$$
\dot \varphi=\sigma_C
$$
and the phase decouples from the system. The shear part
of the dynamical system takes the form \cite{B-H}

\begin{equation}
\dot\sigma_+=-\Theta\sigma_+ -\sigma_A +\frac{1}{3}h^2
(1-\frac{\chi}{2}[(3\gamma-2)\rho+\frac{4}{3}h^2])
\end{equation}

\begin{equation}
\dot\sigma_A=(-\Theta+3\sigma_+ + \sigma_B)\sigma_A
\end{equation}

\begin{equation}
\dot\sigma_B=-\Theta\sigma_B-\sigma^2_A+2\sigma^2_C
\end{equation}

\begin{equation}
\dot\sigma_C=(-\Theta-2\sigma_B)\sigma_C
\end{equation}

Before describing our numerical results for the general
system (1)-(5), (13)-(16) we summarize the conditions for recollaps
in simpler models.
 
   If the matter content of the brane Universe is a perfect
fluid only (so that $h=0$)
, the future recollaps conditions can be 
found analytically. First of all, we can see from eq.(5) that ${\cal U}$
 in this case behaves as  radiation, ${\cal U}={\cal U}_0/a^4$,
 where $a$ is a mean scale factor. The sign of ${\cal U}$ remains unchanged
 during the cosmological evolution.
 
For an isotropic Universe  the recollaps conditions have a very simple form. 
 Indeed, for $\gamma>4/3$
 the ${\cal U}$ - term in eq.(1) falls less rapidly than the matter terms 
making the future recollaps inevitable for any initial negative ${\cal U}$.

 For $\gamma = 4/3$ the first and the last terms
in the r.h.s. of (1) are equally important in a low-energy regime, 
and the condition
for collapse is $\rho< - 6 \chi {\cal U}$. For $2/3 < \gamma < 4/3$ all three
nonzero
terms in r.h.s. of (1) are important. 
Consider, for example, the case of $\gamma=1$.
For the dust-like matter $\rho=\rho_0/a^3$, and assuming that $\rho_0$, 
${\cal U}_0$ are initial values of matter and "dark radiation" densities,
taken when $a_0=1$, we can 
substitut the scale factor 
dependences of $\rho$, $\rho^2$ and ${\cal U}$ into (1). As a result,
we obtain that the
condition $H=0$ leads to a cubic equation for the scale factor

$$
a^3 + 6\chi\frac{{\cal U}_0}{\rho_0}a^2 + \frac{\chi \rho_0}{2} =0.
$$

This equation has positive roots only for $\rho_0^4 < -64 \chi^2 {\cal U}_0^3$.
A root which is bigger than $1$ corresponds to the future of the Universe
under consideration and indicates the 
point of recollaps.
A root with $a<1$ describes the past and corresponds to a bounce. 

A similar treatment can be performed for any other $\gamma>2/3$. 
If $\gamma < 2/3$,
even the quadratic stress-energy correction term, proportional to 
$\rho^2$ falls less rapidly than ${\cal U}$, and the recollaps in
the future becomes impossible. 

If we add a geometrical anisotropy without an anisotropic stress,
the situation does not change significantly. Indeed, the shear scalar
$\Sigma = 3 \sigma_+^2 + \sigma_A^2 + \sigma_B^2 + \sigma_C^2$ behaves
as $\Sigma=\Sigma_0/a^6$, and we only have an additional term in the equation
$H(a)=0$. For example, the system with cosmological constant
$\Lambda$ (which corresponds to $\gamma=0$) and shear gives

$$
\Sigma_0/a^6 + \Lambda + (\chi/2) \Lambda^2 + 6\chi {\cal U}_0/a^4=0
$$
This gives us the condition for $H=0$ to be possible as 
$ \Lambda^2(1+(\chi/2)\Lambda)^2 < -32\Sigma_0\chi^3 {\cal U}_0^3$.

In a similar way we can treat other values of $\gamma$. As  
the shear $\Sigma$ falls more rapidly than the dark radiation ${\cal U}$,
it becomes negligible in the low-energy limit, and 
for $\gamma > 4/3$ with ${\cal U}<0$ a recollaps remains
inevitable. The condition for $\gamma=4/3$ also remaines unchanged.

If we introduce an anisotropic stress, the shear and "dark radiation" can
no longer be expressed in a simple form as a function of the scale 
factor due to source terms in eqs.(5)-(6). In particular, now ${\cal U}$
does {\it not} proportional to $a^{-4}$ and attribution of the name
"dark radiation" to ${\cal U}$ becomes nothing but tradition. 
 The problem of possible future
recollaps of a brane Universe with a homogeneous magnetic field should
be therefore investigated numerically. Our main goal in this paper is to describe
qualitatively the influence of the magnetic field on the possibility
of recollaps, keeping in mind that the non-compactness of our phase space
makes it impossible
to introduce of a simple reasonable measure on it.

Two main features distinguish the magnetic case from the case without
anisotropic stress.
\begin{itemize}
\item The $h^4$-term in (1) has {\it minus} sign, so the recollaps can occur
even with nonnegative dark energy ${\cal U}$ if magnetic field $h$ is large enough.
\item The sigh of dark energy ${\cal U}$ can change during the cosmological 
evolution.
\end{itemize}

The sign change of ${\cal U}$ from positive to negative can 
lead to a recollaps
for positive initial ${\cal U}$ even when $h$ is not large (of course, in
this case ${\cal U}<0$ at the maximal expansion point). The opposite change
from negative to positive values could in principal enlarge zones in 
initial condition space which do not lead to collapse. 
Both these possibilities
are allowed by the equations of motion. 

\begin{figure}
\includegraphics[scale=0.25, angle=0]{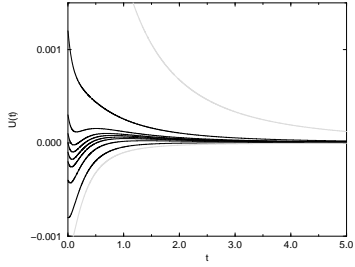}
\caption{Some trajectories which allow a transition from negative 
to positive values of "dark energy"}
\end{figure}

However, in our numerical studies we  have mostly noticed the transition
from {\it plus} to {\it minus} for the sign of ${\cal U}$. Some transitions
{\it minus} $\to$ {\it plus} were constructed using specially adjusted
initial conditions, all of them appeared to be part of a zigzag-like
behavior and are not important for a general picture (see Fig.1). On the
contrary, {\it plus} $\to$ {\it minus} transitions are in some sense typical.
As a result, in all our simulations zones of initial conditions, escaping
a recollaps shrink in comparison with similar
models without a magnetic field.

Some numerical results are shown in Fig. 2 for $\gamma=4/3$ using 
$(H, \mu)$ slices of
initial condition space. In all plots we put $\chi=1$.
White zones represent the initial conditions leading
to eternal expansion, gray zones lead to recollaps, while initial 
conditions from black zones are forbidden by the constraint (1). 
We can see that the
zones, leading to recollaps enlarge and begin to contain some part
of ${\cal U}>0$ initial conditions as compared to $H=0$. However,
there is still a part of initial state space providing an eternal
expansion.
\begin{figure}[h]
\includegraphics[scale=0.2, angle=0]{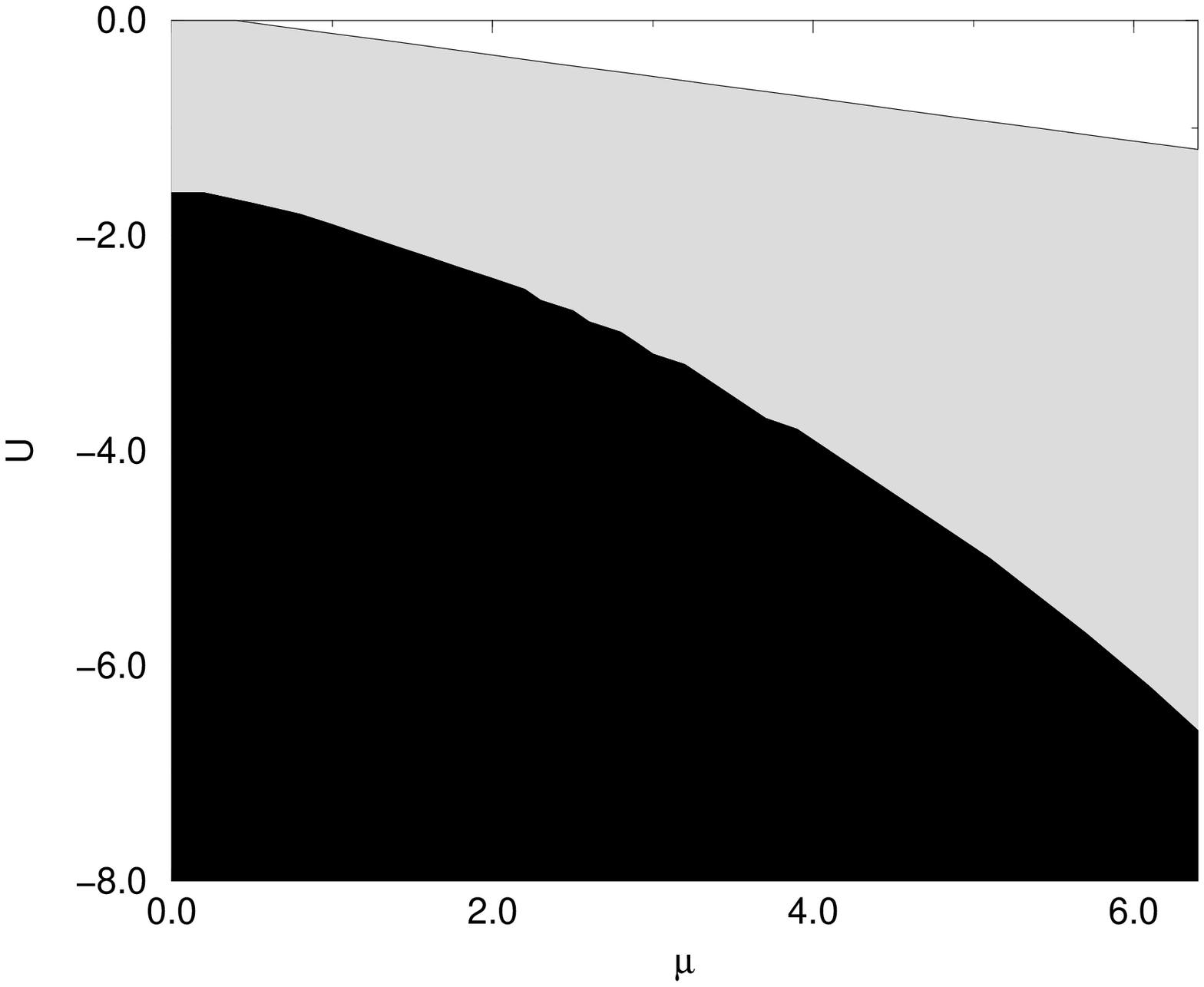}
\includegraphics[scale=0.2, angle=0]{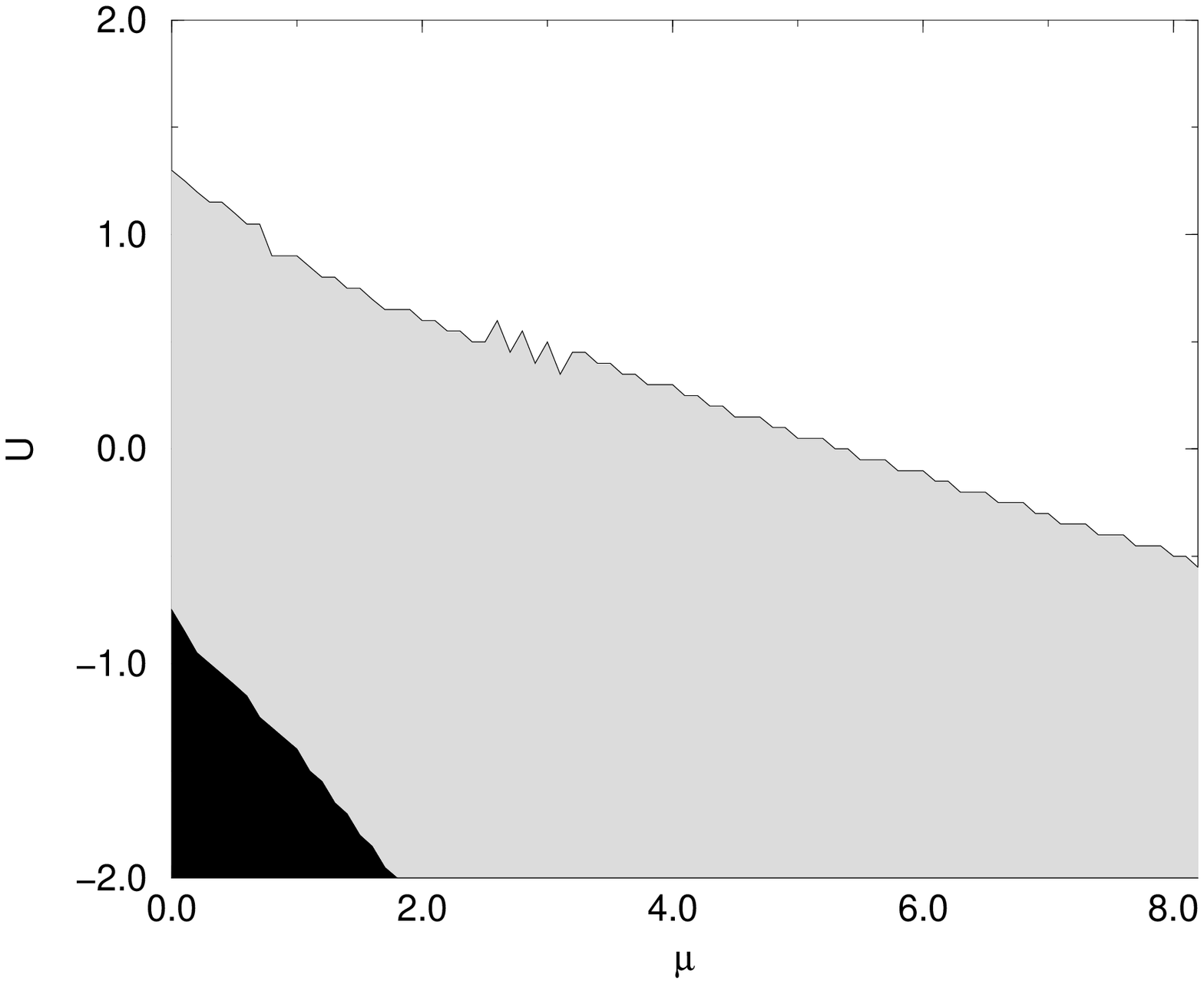}
\caption{The future fate of a brane magnetic Universe, starting
the evolution from $\sigma_A=1.3$, $\sigma_B=\sigma_C=1.1$. 
Initial $h=0$(left) and $h=2$ (right). White 
zone corresponds to eternal expansion, gray zone - to recollaps,
black zone represent initial condition, forbidden by the constraint
equation (1). In this plot $\gamma=4/3$}
\end{figure}

Similar plots for $\gamma=1$ are shown in Fig.3.
\begin{figure}[h]
\includegraphics[scale=0.2,angle=0]{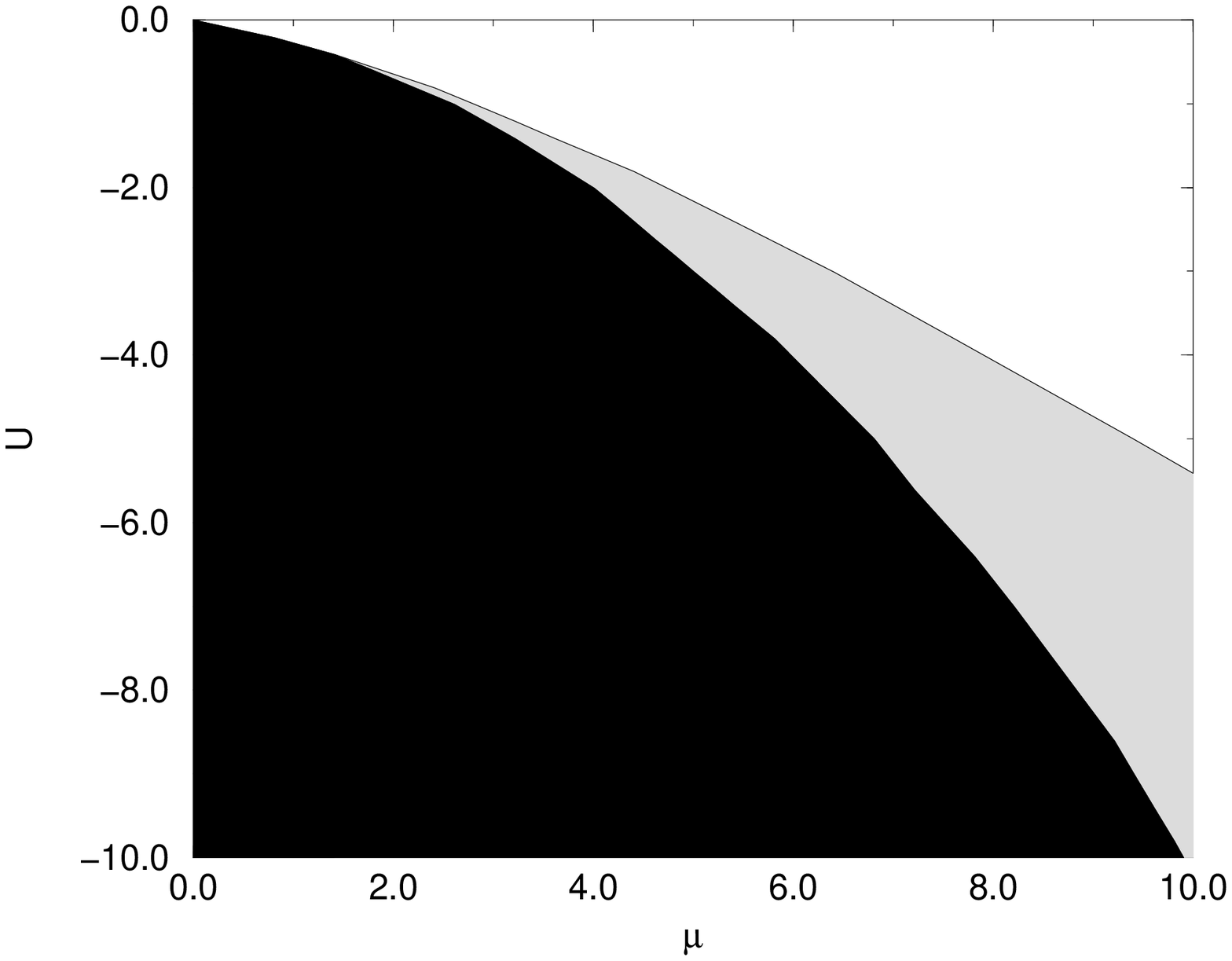}
\includegraphics[scale=0.2,angle=0]{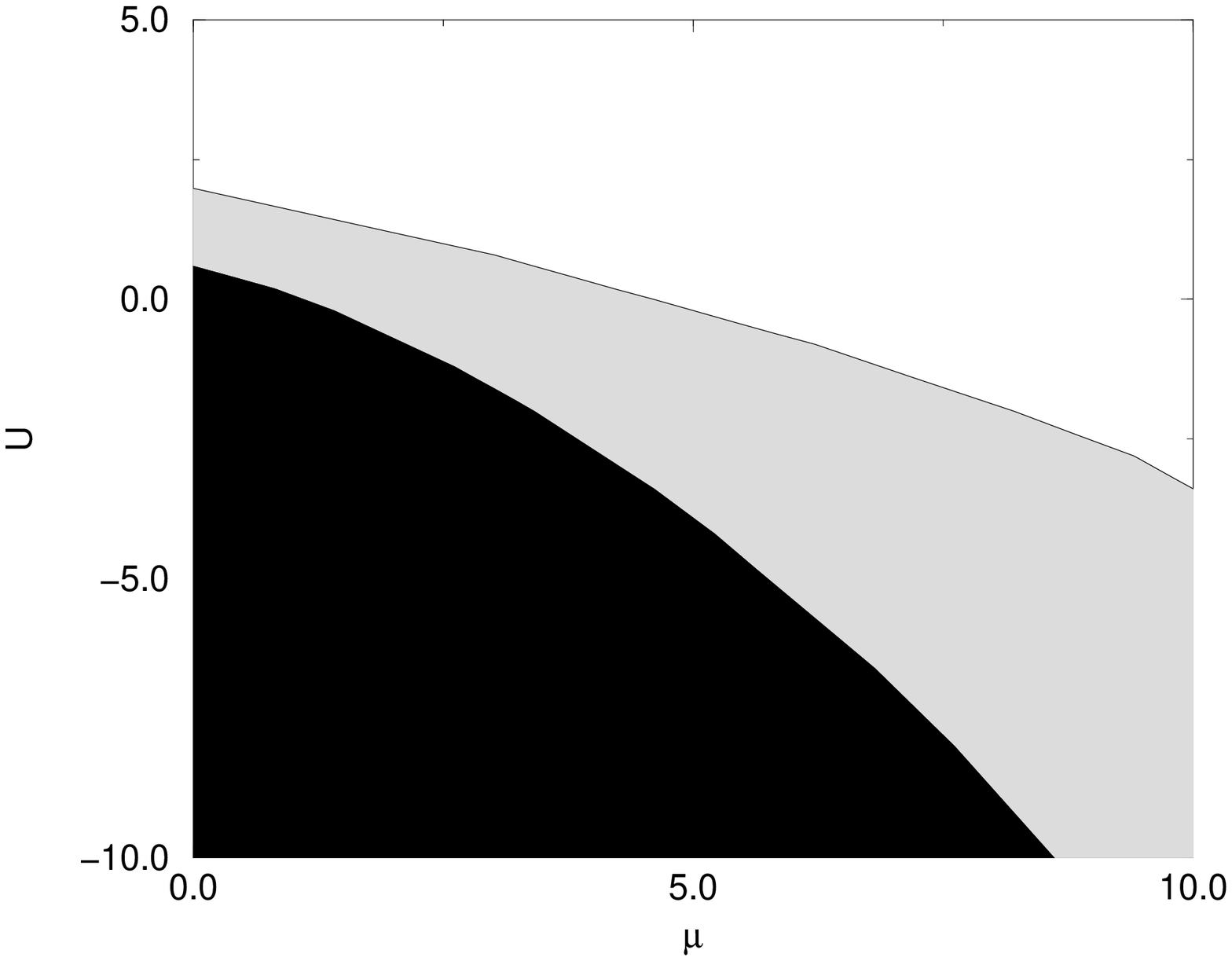}
\caption{The same as in Fig.2, but for $\gamma=1$}
\end{figure}

Different plots, using the $(h, \mu)$ slices of the initial condition space
with the initial ${\cal U}$ fixed
are represented in Fig.4. In particular, we can see how a recollaps
being impossible for initial ${\cal U} \ge 0$ without a magnetic field becomes
possible for nonzero $h$.

As a result, we can see that a homogenios magnetic field, localized
on a brane diminishes the set of initial conditions, which results in
a big Universe, resembling our oun. The set of initial conditions, leading
to a recollaps, increases. However, this change is not drastic, and
we can always find reasonable initial data for an eternally expanding
Universe even if initial magnetic field is large.

\begin{figure}[h]
\includegraphics[scale=0.2, angle=0]{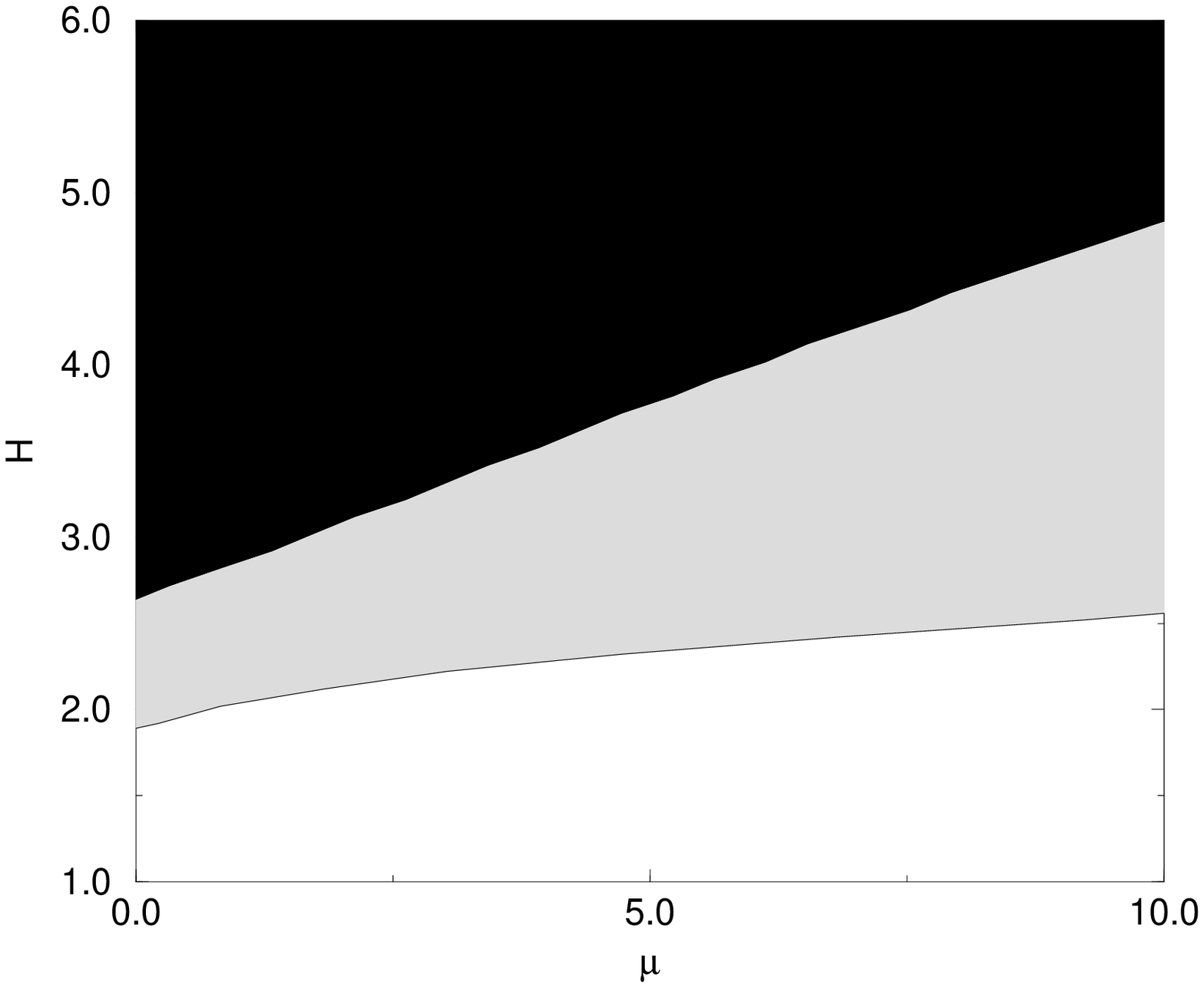}
\includegraphics[scale=0.2, angle=0]{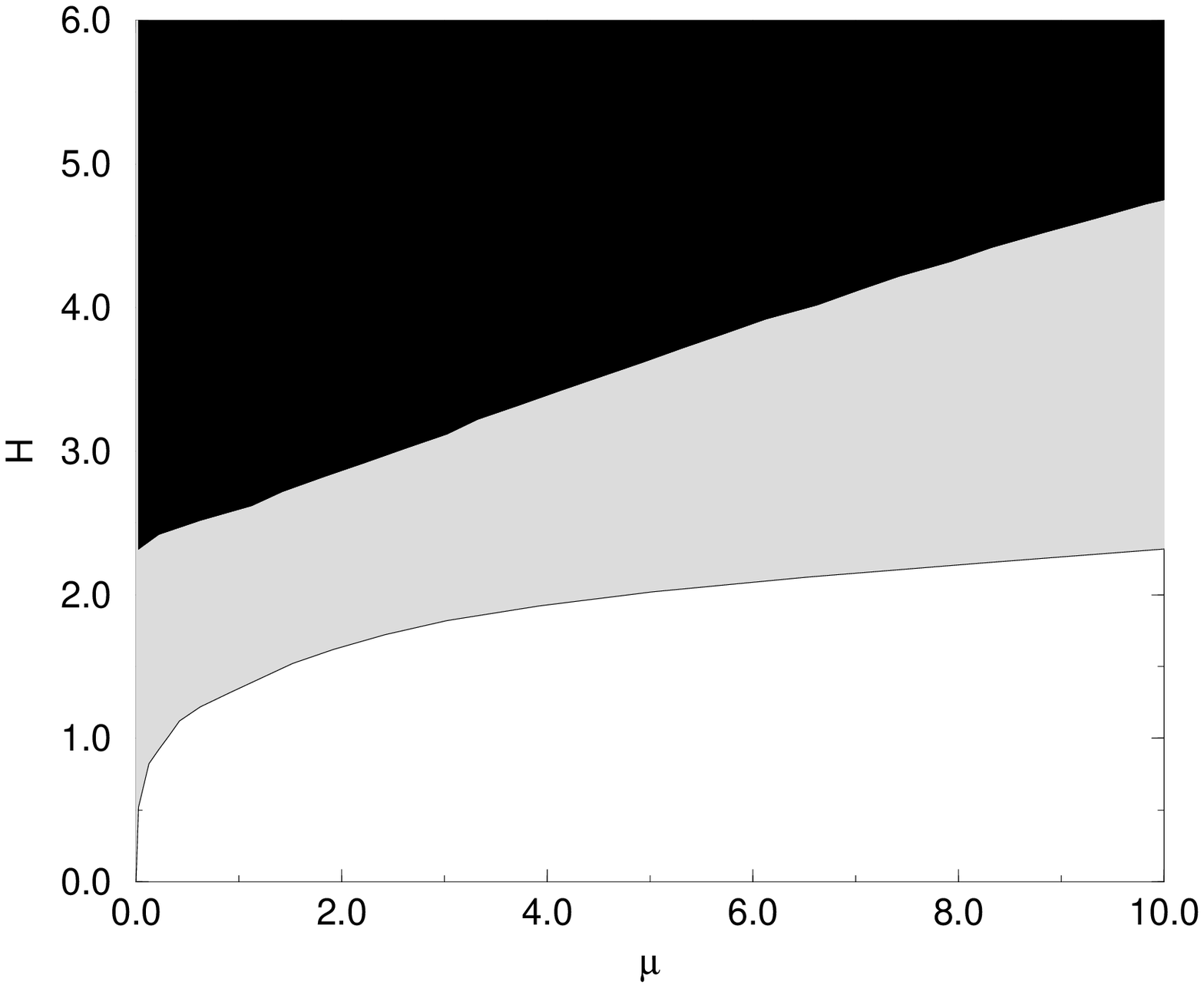}
\includegraphics[scale=0.2, angle=0]{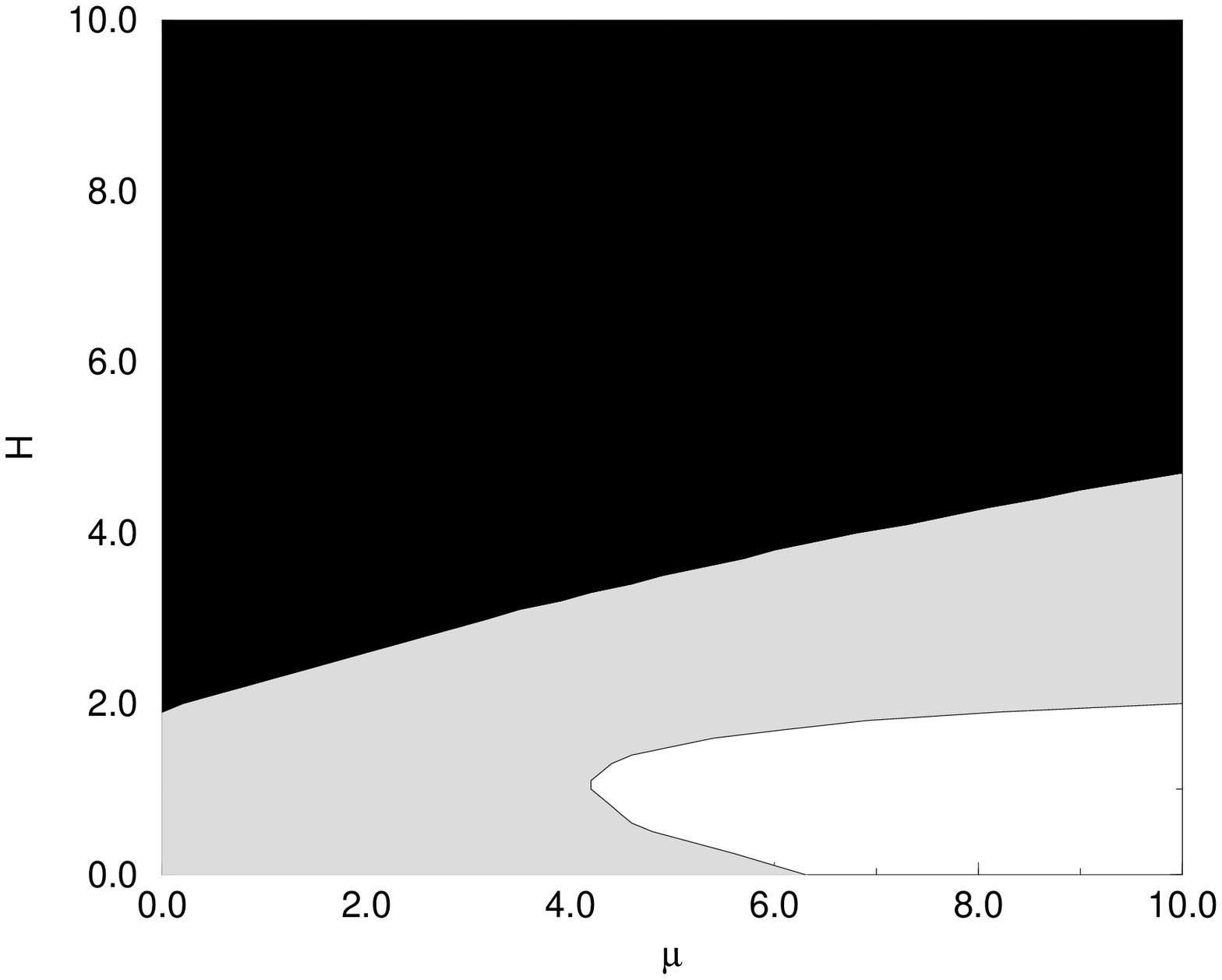}
\includegraphics[scale=0.2, angle=0]{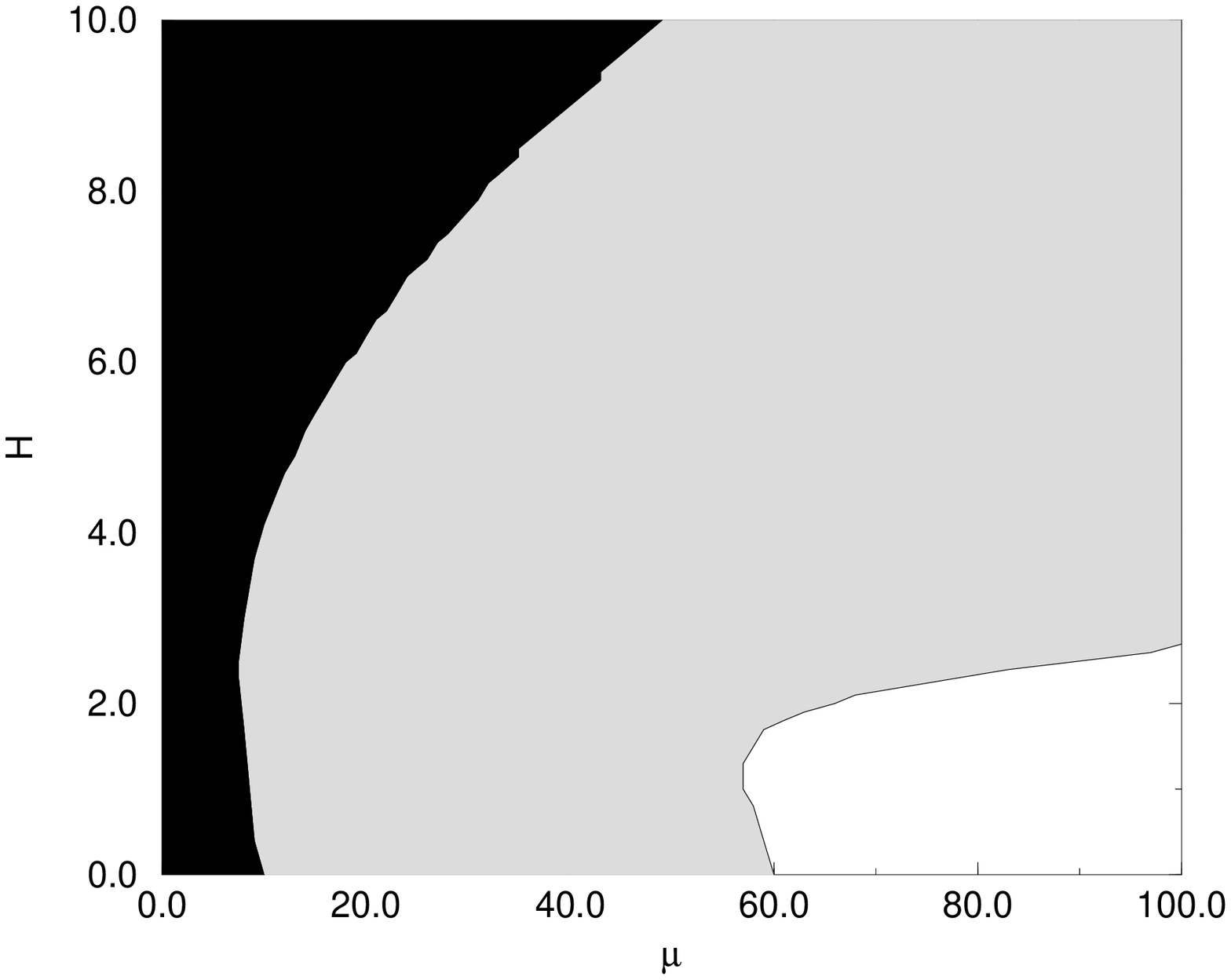}
\caption{The future fate of a brane magnetic Universe,
represented in $(h, \mu)$ slices of the initial condition space. Initial
shear values are the same as in Fig.2. Initial values of "
dark energy" are (from left to the right) ${\cal U}=1$,
${\cal U}=0$, ${\cal U}=-1$ and ${\cal U}=-10$.
The equation of state parameter $\gamma=4/3$.}
\end{figure}

\section*{Acknowledgments}

This work is partially supported by RFBR grant 02-02-16817
and scientific school grant 2338.2003.2 of the Russian Ministry
of Science and Technology. AT is greatful to Sigbjorn Hervik
for discussions.

\end{document}